\documentclass[journal]{IEEEtran}

\usepackage{graphicx} \usepackage{bm} \usepackage{amsmath}
\usepackage{cite}
\usepackage{amssymb} 

\usepackage{amsmath,amsthm,amssymb,mathrsfs}
\usepackage{bm}

\ifCLASSINFOpdf
\else
\fi
\begin{document}

\title{Current-Induced Instability of a Perpendicular Ferromagnet in Spin Hall Geometry}
\author{Tomohiro~Taniguchi$^{1}$,
        Seiji Mitani$^{2}$,
        and Masamitsu~Hayashi$^{2}$ 
        \\
        $^{1}$National Institute of Advanced Industrial Science and Technology (AIST), 
              Spintronics Research Center, 
              Tsukuba, Ibaraki 305-8568, Japan \\
        $^{2}$National Institute for Materials Science, Tsukuba 305-0047, Japan
%\thanks{${}^{\dagger}$Corresponding author. Email address: tomohiro-taniguchi@aist.go.jp}
}

\maketitle

\begin{abstract}
%\boldmath

We develop a theoretical formula of spin Hall torque 
in the presence of two ferromagnets. 
While the direction of the conventional spin Hall torque always points to the in-plane direction, 
the present system enables to manipulate the torque direction acting on one magnetization 
by changing the direction of another magnetization. 
Based on the diffusion equation of the spin accumulation and the Landauer formula, 
we derive analytical formula of the spin Hall torque. 
The present model provides a solution to 
switch a perpendicular ferromagnet deterministically at zero field using the spin Hall effect. 

\end{abstract}


\IEEEpeerreviewmaketitle

% =========================================================================================== %

\section{Introduction}
\label{sec:Introduction}

\IEEEPARstart{S}{pin}-orbit interaction in a nonmagnetic heavy metal 
generates pure spin current flowing along the direction normal to an electric current. 
The phenomenon called spin Hall effect [1-3] has attracted much attention 
as a new method to excite spin transfer torque [4,5] on a magnetization in a ferromagnet layer 
adjacent to the nonmagnetic layer [6-20]. 
In particular, magnetization switching of a perpendicular ferromagnet is an important issue 
for practical application. %such as magnetic random access memory. 

% =========================================================================================== %

Unfortunately however, it is difficult to switch the perpendicular magnetization solely by the spin Hall effect 
due to the following reason. 
Let us assume that an electric current flows along $x$-direction, 
while a ferromagnet is set in $z$-direction. 
Then, the direction of the spin polarization of the pure spin current 
generated by the spin Hall effect is geometrically determined to be induced in $y$-direction [21].
The spin Hall torque exerted by this pure spin current tries to move the magnetization 
from the perpendicular ($z$) direction to the $y$-direction, 
and finally, the magnetization stops its dynamics 
when it becomes parallel to the $y$-axis. 
Since the spin Hall torque does not break the symmetry with respect to the film-plane, 
the magnetization does not cross the film plane. 
Therefore, we cannot switch the magnetization deterministically 
from one equilibrium direction to the other. 
To overcome this problem, usually an in-plane field along the $x$-direction is applied [6]. 
Using a ferromagnet with a tilted anisotropy or non-uniform anisotropy are other solutions [14,16,20]. 

% =========================================================================================== %

% =========================================================================================== %
% =========================================================================================== %

\begin{figure}
  \centerline{\includegraphics[width=1.0\columnwidth]{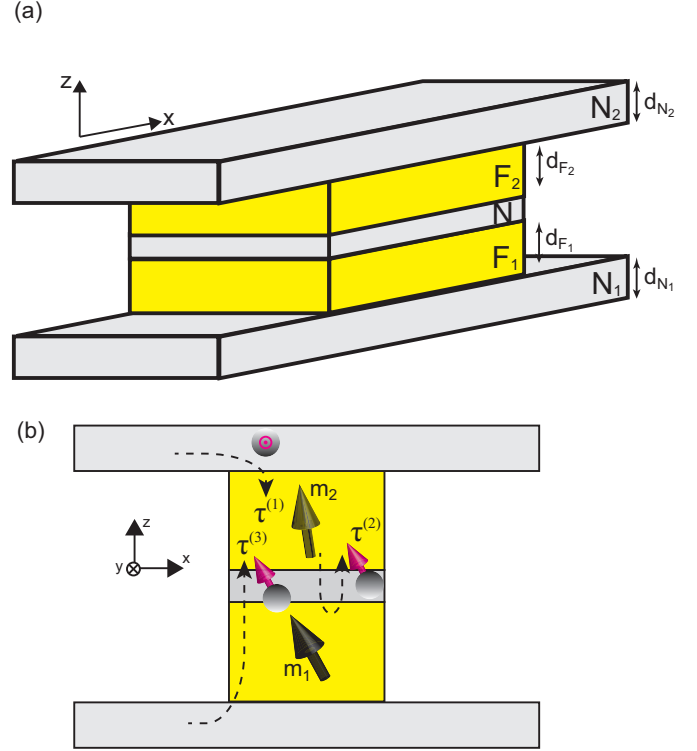}}
  \caption{
           (a) Schematic view of the system consisting of the two ferromagnets %(F${}_{1}$ and F${}_{2}$) 
               and three nonmagnets. %(N, N${}_{1}$, and N${}_{2}$). 
               The thicknesses of the N${}_{k}$ and F${}_{k}$ ($k=1,2$) layers are denoted as 
               $d_{{\rm N}_{k}}$ and $d_{{\rm F}_{k}}$, respectively. 
               The unit vector pointing in the magnetization direction of the F${}_{k}$ layer is $\mathbf{m}_{k}$. %denoted as $\mathbf{m}_{k}$. 
           (b) The spin Hall torques acting on the magnetization of the F${}_{2}$ layer are schematically shown. 
  \vspace{-3.0ex}}
  \label{fig:fig1}
\end{figure}

% =========================================================================================== %
% =========================================================================================== %

In this paper, we propose another solution to switch perpendicular ferromagnet deterministically 
by using the spin Hall effect. 
As schematically shown in Fig. \ref{fig:fig1}(a), 
the system consists of two ferromagnets, F${}_{1}$ and F${}_{2}$, 
and three nonmagnets, N, N${}_{1}$, and N${}_{2}$. 
The two nonmagnets, N${}_{1}$ and N${}_{2}$, show spin-orbit interactions, %strong spin-orbit interactions, 
and inject pure spin current into the F${}_{1}$ and F${}_{2}$ layers via the spin Hall effect. 
The N layer is just a spacer, 
and does not show spin-orbit interaction. 
%Instead, the spin diffusion length of the N layer is sufficiently long 
%compared with its thickness, 
%and therefore, the spin diffusion in the N layer is negligible. 
Current is passed along the $x$-direction and flows within the N${}_{1}$ and N${}_{2}$ layers. 
%Current flows within the N${}_{1}$ and N${}_{2}$ layers along the $x$-direction. 
The F${}_{2}$ layer is a free layer, 
and has a perpendicular anisotropy. 
The spin Hall torque acting on the magnetization of the F${}_{2}$ layer 
consists of three contributions. 
The first one is the conventional spin Hall torque caused by the pure spin current injected from the N${}_{2}$ layer, 
which is denoted as $\bm{\tau}^{(1)}$ in Fig. \ref{fig:fig1}(b). 
The second contribution comes from the electrons 
passing through from N${}_{2}$ to N${}_{1}$ layer. 
Due to the diffusion between the N${}_{1}$ and N${}_{2}$ layer, 
these electrons can return back to the F${}_{2}$ layer, and excites the spin Hall torque denoted as $\bm{\tau}^{(2)}$ in Fig. \ref{fig:fig1}(b). 
The third contribution comes from the pure spin current originated in the N${}_{1}$ layer, 
and is denoted as $\bm{\tau}^{(3)}$ in Fig. \ref{fig:fig1}(b). 
While the first torque $\bm{\tau}^{(1)}$ points to the $y$-direction, as in the case of the conventional spin Hall system, 
the directions of $\bm{\tau}^{(2)}$ and $\bm{\tau}^{(3)}$ depend on 
the magnetization of the F${}_{1}$ layer. 
This is because the longitudinal and transverse spin currents [22-28] have different relaxation length scales; 
here the longitudinal and transverse correspond to parallel and perpendicular to the local magnetization, respectively. 
%where the longitudinal and transverse means parallel and perpendicular to the local magnetization. 
Deterministic switching of the F${}_{2}$ layer is possible 
when the F${}_{1}$ layer tilts from the $z$-axis. 
%Then, if the magnetization in the F${}_{1}$ layer tilts from the $z$-axis and breaks the symmetry, 
%the magnetization in the F${}_{2}$ layer can be deterministically switched 
%due to the torques, $\bm{\tau}^{(2)}$ and $\bm{\tau}^{(3)}$. 
In the following, we derive the explicit forms of the spin Hall torques, $\bm{\tau}^{(i)}$ ($i=1,2,3$), 
based on the diffusion equation of the spin accumulation and the Landauer formula. 

% =========================================================================================== %

This paper is organized as follows. 
In Sec. \ref{sec:Spin Hall Torque Formula}, we derive a theoretical formula of the spin torque 
acting on the magnetization of the F${}_{2}$ layer in the geometry given by Fig. \ref{fig:fig1}(a). 
In Sec. \ref{sec:LLG Equation}, we show evidence of the deterministic switching in the F${}_{2}$ layer 
by solving the Landau-Lifshitz-Gilbert (LLG) equation. 
The conclusion is summarized in Sec. \ref{sec:Conclusion}.

% =========================================================================================== %

% =========================================================================================== %

\section{Spin Hall Torque Formula}
\label{sec:Spin Hall Torque Formula}

The system we consider is schematically shown in Figs. \ref{fig:fig1}(a) and \ref{fig:fig1}(b). 
We denote the thicknesses of the N${}_{k}$ and F${}_{k}$ layers as $d_{{\rm N}_{k}}$ and $d_{{\rm F}_{k}}$, respectively ($k=1,2$). 
The unit vector pointing in the magnetization direction of the F${}_{k}$ layer is 
denoted as $\mathbf{m}_{k}$. 
The electric field $E_{x}$ is applied along the $x$-direction, 
and generates an electric current density in the nonmagnet [21] %given by [21] 
\begin{equation}
  J_{e}
  =
  \sigma_{{\rm N}_{k}}
  E_{x}
  +
  \frac{\vartheta_{{\rm N}_{k}} \sigma_{{\rm N}_{k}}}{2e}
  \mathbf{e}_{y}
  \cdot
  \partial_{z}
  \bm{\mu}_{{\rm N}_{k}},
  \label{eq:electric_current}
\end{equation}
where $\sigma_{{\rm N}_{k}}$ and $\vartheta_{{\rm N}_{k}}$ are
the conductivity and the spin Hall angle of the N${}_{k}$ layer, respectively. 
The spin accumulation in the N${}_{k}$ layer is denoted as $\bm{\mu}_{{\rm N}_{k}}$. 
Similarly, the spin current density in the nonmagnet flowing along the $z$-direction is [21] %given by [21] 
\begin{equation}
  \mathbf{J}_{s}
  =
  -\frac{\hbar \sigma_{{\rm N}_{k}}}{4e^{2}}
  \partial_{z}
  \bm{\mu}_{{\rm N}_{k}}
  +
  \frac{\hbar \vartheta_{{\rm N}_{k}}}{2e}
  E_{x}
  \mathbf{e}_{y},
  \label{eq:spin_current}
\end{equation}
where the vector represents the direction of the spin polarization. 
The spin accumulation in the nonmagnet obeys the diffusion equation, %nonmagnetic layer obeys the diffusion equation, 
%$\partial^{2}\bm{\mu}_{\rm N}/\partial z^{2}=\bm{\mu}_{\rm N}/\lambda_{\rm N}^{2}$, 
and its solution can be expressed by a linear combination of $e^{\pm z/\lambda_{\rm N}}$, 
where $\lambda_{\rm N}$ is the spin diffusion length. %of the nonmagnet. 

% =========================================================================================== %

The pure spin current injected from the nonmagnet creates a spin accumulation in the ferromagnet. 
The spin accumulation in a ferromagnet, $\bm{\mu}_{\rm F}$, 
can be decomposed into the longitudinal and transverse components as 
$\bm{\mu}_{\rm F}^{\rm L}=(\mathbf{m}\cdot\bm{\mu}_{\rm F})\mathbf{m}$ and 
$\bm{\mu}_{\rm F}^{\rm T}=\mathbf{m}\times(\bm{\mu}_{\rm F}\times\mathbf{m})=\bm{\mu}_{\rm F}-\bm{\mu}_{\rm F}^{\rm L}$, respectively. 
The relaxation length scale of the longitudinal spin accumulation, called the spin diffusion length, 
depends on the spin-flip scattering time, and is on the order of 1-10 nm [22]. 
The relaxation length of the transverse spin accumulation, called the penetration depth of the transverse spin current, 
depends on both the spin-flip scattering time and precession period of spin around the local magnetization due to the exchange coupling. 
Since the penetration depth of the transverse spin current is usually shorter than the spin diffusion length [23-28], 
let us assume that the spin accumulation in the F layer, $\bm{\mu}_{\rm F}$, has only the longitudinal component, for simplicity. 
In other words, the direction of the spin polarization of the electrons becomes 
parallel to the magnetization direction when it is injected into the ferromagnet. 
We note that a finite penetration depth of the transverse spin current can be taken into account in the following calculations, 
as done in the case of spin pumping [26]. 
It does not however affect the main conclusion significantly, in spite of the fact that it makes the calculations complex. 
The important point is that the longitudinal and transverse spin currents relax with different length scales, 
and therefore, the direction of the spin polarization of the spin current is modified 
by passing through a ferromagnet. 
The idea is similar to what is used in Ref. [29]. 

% =========================================================================================== %

%The boundary conditions of the diffusion equation are as follows. 
At both ends of the nonmagnets, N${}_{1}$ and N${}_{2}$, 
the spin currents are zero. 
On the other hand, the spin current density at the F/N interface 
flowing from F to N layer is given [23] %by the Landauer formula [23], 
\begin{equation}
\begin{split}
  \mathbf{J}_{s}^{\rm F \to N}
  =&
  \frac{1}{4\pi S}
  \left[
    \frac{(1-\gamma_{\rm F/N}^{2})g_{\rm F/N}}{2}
    \mathbf{m}
    \cdot
    \left(
      \bm{\mu}_{\rm F}
      -
      \bm{\mu}_{\rm N}
    \right)
    \mathbf{m}
  \right.
\\
  &
    -
  \left.
    g_{\rm r(F/N)}
    \mathbf{m}
    \times
    \left(
      \bm{\mu}_{\rm N}
      \times
      \mathbf{m}
    \right)
  \right],
  \label{eq:spin_current_FN}
\end{split}
\end{equation}
where $S$ is the cross section area of the $xy$-plane 
(Eq. (\ref{eq:spin_current_FN}) is applicable to both F${}_{k}$/N and F${}_{k}$/N${}_{k}$ interfaces). 
Here $g=g^{\uparrow}+g^{\downarrow}$ and $\gamma=(g^{\uparrow}-g^{\downarrow})/(g^{\uparrow}+g^{\downarrow})$ 
are the total F/N interface conductance %, i.e., the sum of the conductances of spin-up and spin-down electrons, 
and its spin polarization, respectively. 
We neglect spin-flip scattering at the interface. 
The conductance $g$ is related to the F/N interface resistance $r$ via $g/S=h/(e^{2}r)$. 
The real part of the mixing conductance is denoted as $g_{\rm r}$. 
For simplicity, we neglect the imaginary part of the mixing conductance, 
which is usually negligible in a metallic current-perpendicular-to-plane system [23,24]
but might be large in spin Hall geometry [11-16]. 
The spin Hall torque acting on the magnetization of the F${}_{k}$ layer %originates from the absorption of the transverse spin current 
%at the F${}_{k}$/N${}_{k}$ and F${}_{k}$/N interfaces, and therefore, 
is given by 
\begin{equation}
  \frac{{\rm d}\mathbf{m}_{k}}{{\rm d}t}
  \bigg|_{\rm STT}
  =
  \frac{\gamma_{0}}{M_{k}Sd_{{\rm F}_{k}}}
  \mathbf{m}_{k}
  \times
  \left[
    \left(
      \mathbf{J}_{s}^{{\rm F}_{k} \to {\rm N}_{k}}
      +
      \mathbf{J}_{s}^{{\rm F}_{k} \to {\rm N}}
    \right)
    \times
    \mathbf{m}_{k}
  \right],
  \label{eq:STT_def}
\end{equation}
where $\gamma_{0}$ is the gyromagnetic ratio of the F${}_{k}$ layer. 

% =========================================================================================== %

We assume that the N layer sandwiched by the two ferromagnets 
is sufficiently thin compared to its spin diffusion length, 
and therefore, the spin current in the N layer is conserved, 
i.e., $\mathbf{J}_{s}^{{\rm F}_{1} \to {\rm N}} + \mathbf{J}_{s}^{{\rm F}_{2} \to {\rm N}}=\bm{0}$. 
Solving the diffusion equations of the N${}_{k}$ and F${}_{k}$ layers, 
then substituting the solutions to Eq. (\ref{eq:spin_current_FN}), 
and using the relation $\mathbf{J}_{s}^{{\rm F}_{1} \to {\rm N}} + \mathbf{J}_{s}^{{\rm F}_{2} \to {\rm N}}=\bm{0}$, 
Eq. (\ref{eq:STT_def}) becomes 
\begin{equation}
\begin{split}
  \frac{{\rm d}\mathbf{m}_{k}}{{\rm d}t}
  \bigg|_{\rm STT}
  =&
  (-1)^{k}
  \tau_{k}^{(1)}
  \mathbf{m}_{k}
  \times
  \left(
    \mathbf{e}_{y}
    \times
    \mathbf{m}_{k}
  \right)
\\
  &
  -
  (-)^{k+1}
  \left[
    \tau_{k}^{(2)}
    m_{k y}
    \mathbf{m}_{1}
    \cdot
    \mathbf{m}_{2}
    -
    \tau_{k}^{(3)}
    m_{\ell y}
  \right]
\\
  &
  \ \ \ \ \ \ \ \times
  \frac{\mathbf{m}_{k} \times (\mathbf{m}_{\ell} \times \mathbf{m}_{k})}{1-\lambda_{1}^{\prime}\lambda_{2}^{\prime}(\mathbf{m}_{1}\cdot\mathbf{m}_{2})^{2}},
  \label{eq:LLG_STT}
\end{split}
\end{equation}
where $(k,\ell)=(1,2)$ or $(2,1)$. 
%We introduce the following notations for simplicity; 
The following notations are introduced;
\begin{equation}
  \tau_{k}^{(1)}
  =
  \frac{\gamma_{0} \hbar \vartheta_{{\rm N}_{k}} g_{{\rm r}({\rm F}_{k}/{\rm N}_{k})}^{\prime} \sigma_{{\rm N}_{k}} E_{x}}{2eg_{{\rm sd}({\rm N}_{k})}M_{k}d_{{\rm F}_{k}}}
  \tanh
  \left(
    \frac{d_{{\rm N}_{k}}}{2 \lambda_{{\rm N}_{k}}}
  \right),
\end{equation}
\begin{equation}
  \tau_{k}^{(2)}
  =
  \lambda_{\ell}^{\prime}
  \frac{\gamma_{0} \hbar \vartheta_{{\rm N}_{k}} g_{{\rm r}({\rm F}_{k}/{\rm N})} g_{{\rm F}_{k}/{\rm N}}^{\prime} \sigma_{{\rm N}_{k}} E_{x}}
    {2e \tilde{g}_{k} [g_{{\rm r}({\rm F}_{\ell}/{\rm N})}+g_{{\rm F}_{k}/{\rm N}}^{\prime}]M_{k} d_{{\rm F}_{k}}},
\end{equation}
\begin{equation}
  \tau_{k}^{(3)}
  =
  \frac{\gamma_{0} \hbar \vartheta_{{\rm N}_{\ell}} g_{{\rm r}({\rm F}_{k}/{\rm N})} g_{{\rm F}_{\ell}/{\rm N}}^{\prime} \sigma_{{\rm N}_{\ell}} E_{x}}
    {2e \tilde{g}_{\ell} [g_{{\rm r}({\rm F}_{k}/{\rm N})}+g_{{\rm F}_{\ell}/{\rm N}}^{\prime}]M_{k} d_{{\rm F}_{k}}},
\end{equation}
\begin{equation}
  \lambda_{k}^{\prime}
  =
  \frac{g_{{\rm r}({\rm F}_{k}/{\rm N})} - g_{{\rm F}_{k}/{\rm N}}^{\prime}}
    {g_{{\rm r}({\rm F}_{\ell}/{\rm N})} + g_{{\rm F}_{k}/{\rm N}}^{\prime}},
\end{equation}
where 
\begin{equation}
  \frac{1}{g_{{\rm r}({\rm F}_{k}/{\rm N}_{k})}^{\prime}}
  =
  \frac{1}{g_{{\rm r}({\rm F}_{k}/{\rm N}_{k})}}
  +
  \frac{1}{g_{{\rm sd}({\rm N}_{k})} \tanh(d_{{\rm N}_{k}}/\lambda_{{\rm N}_{k}})},
\end{equation}
\begin{equation}
\begin{split}
  \frac{1}{g_{{\rm F}_{k}/{\rm N}_{k}}^{\prime}}
  =&
  \frac{2}{(1-\gamma_{{\rm F}_{k}/{\rm N}_{k}}^{2})g_{{\rm F}_{k}/{\rm N}_{k}}}
  +
  \frac{1}{g_{{\rm sd}({\rm F}_{k})} \tanh(d_{{\rm F}_{k}}/\lambda_{{\rm F}_{k}})}
\\
  &+
  \frac{1}{g_{{\rm sd}({\rm N}_{k})} \tanh(d_{{\rm N}_{k}}/\lambda_{{\rm N}_{k}})},
\end{split}
\end{equation}
\begin{equation}
\begin{split}
  \frac{1}{g_{{\rm F}_{k}/{\rm N}}^{\prime}}
  =&
  \frac{2}{(1-\gamma_{{\rm F}_{k}/{\rm N}}^{2})g_{{\rm F}_{k}/{\rm N}}}
  +
  \frac{1}{g_{{\rm sd}({\rm F}_{k})} \tanh(d_{{\rm F}_{k}}/\lambda_{{\rm F}_{k}})}
\\
  &-
  \frac{g_{{\rm F}_{k}/{\rm N}_{k}}^{\prime}}{[ g_{{\rm sd}({\rm F}_{k})} \sinh(d_{{\rm F}_{k}}/\lambda_{{\rm F}_{k}}) ]^{2}},
\end{split}
\end{equation}
\begin{equation}
\begin{split}
  \frac{1}{\tilde{g}_{k}}
  =
  \frac{g_{{\rm F}_{k}/{\rm N}_{k}}^{\prime} \tanh[d_{{\rm N}_{k}}/(2 \lambda_{{\rm N}_{k}})]}
    {g_{{\rm sd}({\rm F}_{k})} g_{{\rm sd}({\rm N}_{k})} \sinh(d_{{\rm F}_{k}}/\lambda_{{\rm F}_{k}})}, 
\end{split}
\end{equation}
\begin{align}
 g_{{\rm sd}({\rm F}_{k})}
 =
 \frac{h (1-\beta_{{\rm F}_{k}}^{2}) S}{2e^{2} \rho_{{\rm F}_{k}} \lambda_{{\rm F}_{k}}},
&&
 g_{{\rm sd}({\rm N}_{k})}
 =
 \frac{h S}{2e^{2} \rho_{{\rm N}_{k}} \lambda_{{\rm N}_{k}}},
\end{align}
%\begin{equation}
% g_{{\rm sd}({\rm F}_{k})}
% =
% \frac{h (1-\beta_{{\rm F}_{k}}^{2}) S}{2e^{2} \rho_{{\rm F}_{k}} \lambda_{{\rm F}_{k}}},
%\end{equation}
%\begin{equation}
% g_{{\rm sd}({\rm N}_{k})}
% =
% \frac{h S}{2e^{2} \rho_{{\rm N}_{k}} \lambda_{{\rm N}_{k}}},
%\end{equation}
and $\rho=1/\sigma$. 
The spin polarization of the conductivity is denoted as $\beta=(\sigma^{\uparrow}-\sigma^{\downarrow})/(\sigma^{\uparrow}+\sigma^{\downarrow})$. 
%A similar calculation was performed in Ref. [29]. 
A calculation of spin Hall torque in a system including two ferromagnets 
with a different geometry was developed in Ref. [30]. 
%Adding the spin pumping effect will be another interesting subject [35]. 

% =========================================================================================== %

%The physical meanings of the three torques appeared on the right-hand side of Eq. (\ref{eq:LLG_STT}) have been briefly described in Sec. \ref{sec:Introduction}. 
In Eq. (\ref{eq:LLG_STT}), 
the first torque proportional to $\tau_{k}^{(1)}$ originates from 
the pure spin current injected from the N${}_{k}$ layer via the spin Hall effect, 
and points to the $y$-direction, as in the case of the conventional spin Hall torque. 
The second term proportional to $\tau_{k}^{(2)}$ originates from 
the electrons passing from the N${}_{k}$ layer through the F${}_{k}$ layer 
and being reflected from the N${}_{\ell}$ layer ($\ell \neq k$)
%the diffusion of the spin current injected from the N${}_{k}$ layer. 
This torque becomes zero when $m_{ky}=0$ 
because the spin current from the N${}_{k}$ layer is 
completely absorbed at the F${}_{k}$/N${}_{k}$ interface, %due to the absorption of the transverse spin current, 
and therefore the electrons passing through the F${}_{k}$ layer 
and diffusing between the N${}_{1}$ and N${}_{2}$ layers do not have net spin polarization. 
Similarly, this torque also becomes zero when $\mathbf{m}_{1}\cdot\mathbf{m}_{2}=0$ 
because in this case the spin current is completely absorbed 
at the F${}_{k}$/N${}_{k}$ or F${}_{\ell}$/N interface. 
%This torque becomes zero when $m_{ky}=0$ or $\mathbf{m}_{1}\cdot\mathbf{m}_{2}=0$, 
%because in these cases the spin polarization has only the transverse components 
%with respect to $\mathbf{m}_{1}$ or $\mathbf{m}_{2}$, 
%and therefore, the net spin polarization of the spin current diffused between 
%the F${}_{1}$ and F${}_{2}$ layer is zero. 
The third term proportional to $\tau_{k}^{(3)}$ originates from 
the pure spin current injected from the N${}_{\ell}$ layer. 
Due to the absorption of the transverse spin current in the N${}_{\ell}$ layer, 
this torque is zero when $m_{\ell y}=0$. 
%The F${}_{k}$ layer absorbs the transverse component of the spin current, 
%and thus, only the longitudinal component of the spin current survives during the transport 
%from the N${}_{k}$ layer to the F${}_{k}$ layer. 
%If $m_{ky}=0$, the spin current injected from the N${}_{k}$ layer has only the transverse component, 
%and therefore, the electrons passing through the F${}_{k}$ layer does not have net spin polarization. 
%Therefore, even if these electrons backs to the F${}_{k}$ layer due to the diffusion, 
%they do not contribute to the spin Hall torque, and therefore, $\tau_{k}^{(2)}=0$. 
%Similarly, when $\mathbf{m}_{1}\cdot\mathbf{m}_{2}=0$, 
%the spin polarization of the spin current passing through the F${}_{k}$ layer is completely absorbed 
%by the F${}_{\ell}$ layer, 
%and therefore, it does not contribute to $\tau_{k}^{(2)}$. 

% =========================================================================================== %

%We emphasize that the directions of the second and third torques acting on the F${}_{k}$ layer, $\mathbf{m}_{k}\times(\mathbf{m}_{\ell}\times\mathbf{m}_{k})$, 
The directions of the second and third torques acting on the F${}_{k}$ layer, $\mathbf{m}_{k}\times(\mathbf{m}_{\ell}\times\mathbf{m}_{k})$, 
can be changed by controlling the magnetization direction of the other layer, $\mathbf{m}_{\ell}$. 
This is because the direction of the spin polarization is modified by passing through the ferromagnet 
due to the different relaxation lengths between the longitudinal and transverse spin currents. 
Now let us assume that the magnetization in the F${}_{1}$ layer, $\mathbf{m}_{1}$, tilts from the $z$-axis. 
%as schematically shown in Fig. \ref{fig:fig1}(b). 
Then the second and third torques acting on the magnetization of the F${}_{2}$ layer 
remain finite even when $\mathbf{m}_{2}$ arrives at the film-plane, 
contrary to the first torque which becomes zero when $\mathbf{m}_{2}$ becomes parallel to the $y$-axis, 
as in the case of the conventional spin Hall torque. 
Depending on the direction of the electric field $E_{x}$, 
the second and third torques move the magnetization $\mathbf{m}_{2}$ to the positive or negative $z$-direction. 
Therefore, a deterministic switching of $\mathbf{m}_{2}$ can be expected. 
%We note that a finite tilted angle of $\mathbf{m}_{1}$ is necessary 
A finite tilted angle of $\mathbf{m}_{1}$ is necessary 
to break the symmetry with respect to the film-plane 
and induces the deterministic switching. 
If $\mathbf{m}_{1}\parallel\mathbf{e}_{z}$, 
and when $\mathbf{m}_{2}$ arrives at the film-plane, 
the second and third torques in Eq. (\ref{eq:LLG_STT}) become zero. 
Then, the magnetization will be stopped at the plane, as in the case of the conventional spin Hall system. 

% =========================================================================================== %

\section{LLG Equation}
\label{sec:LLG Equation}

Using Eq. (\ref{eq:LLG_STT}), 
the LLG equation of the F${}_{k}$ layer becomes
\begin{equation}
\begin{split}
  \frac{{\rm d}\mathbf{m}_{k}}{{\rm d}t}
  =&
  -\gamma_{0}
  \mathbf{m}_{k}
  \times
  \mathbf{H}_{k}
  +
  \bm{\tau}_{k}
  +
  \alpha
  \mathbf{m}_{k}
  \times
  \frac{{\rm d}\mathbf{m}_{k}}{{\rm d}t},
%\\
%  &
%  +
%  \bm{\tau}_{k}^{(1)}
%  +
%  \bm{\tau}_{k}^{(2)}
%  +
%  \bm{\tau}_{k}^{(3)},
  \label{eq:LLG}
\end{split}
\end{equation}
where 
the spin Hall torque $\bm{\tau}_{k}$ is given by Eq. (\ref{eq:LLG_STT}). 
%\begin{equation}
%  \bm{\tau}_{k}^{(1)}
%  =
%  (-1)^{k}
%  \tau_{k}^{(1)}
%  \mathbf{m}_{k}
%  \times
%  \left(
%    \mathbf{e}_{y}
%    \times
%    \mathbf{m}_{k}
%  \right), 
%\end{equation}
%\begin{equation}
%  \bm{\tau}_{k}^{(2)}
%  =
%  (-1)^{k}
%  \tau_{k}^{(2)}
%  m_{ky}
%  \mathbf{m}_{1}\cdot\mathbf{m}_{2} 
%  \frac{\mathbf{m}_{k} \times (\mathbf{m}_{\ell}\times\mathbf{m}_{k})}{1-\lambda_{1}^{\prime}\lambda_{2}^{\prime}(\mathbf{m}_{1}\cdot\mathbf{m}_{2})^{2}},
%\end{equation}
%\begin{equation}
%  \bm{\tau}_{k}^{(3)}
%  =
%  (-1)^{k+1}
%  \tau_{k}^{(3)}
%  m_{\ell y}
%  \frac{\mathbf{m}_{k} \times (\mathbf{m}_{\ell}\times\mathbf{m}_{k})}{1-\lambda_{1}^{\prime}\lambda_{2}^{\prime}(\mathbf{m}_{1}\cdot\mathbf{m}_{2})^{2}}. 
%\end{equation}
As mentioned above, we assume that the magnetization of the F${}_{1}$ layer is fixed, 
while that of the F${}_{2}$ layer changes its direction according to Eq. (\ref{eq:LLG}). 
The magnetic field acting on $\mathbf{m}_{2}$, $\mathbf{H}_{2}=H_{\rm K}m_{2z} \mathbf{e}_{z}$, 
consists of the perpendicular anisotropy field $H_{\rm K}$. 
The Gilbert damping constant of the F${}_{2}$ layer is denoted as $\alpha$. %is denoted as $\alpha$. 
In the absence of the electric field $E_{x}$, 
the F${}_{2}$ layer has two stable states, $\mathbf{m}_{2}=\pm\mathbf{e}_{z}$. 
In this calculation, we assume that the magnetization $\mathbf{m}_{2}$ initially points to the positive $z$-direction. 

% =========================================================================================== %

%For simplicity, a symmetric system is adopted, 
%i.e., the parameters of the N${}_{1}$ and N${}_{2}$ layers, 
We assume that the parameters of the N${}_{1}$ and N${}_{2}$ layers, 
as well as those of the F${}_{1}$ and F${}_{2}$ layers, are identical. 
We define the current density $j$ as $j=\sigma_{{\rm N}_{1}}E_{x}=\sigma_{{\rm N}_{2}}E_{x}$. 
Linearizing Eq. (\ref{eq:LLG}) around the initial state, 
we found that the magnetization of the free (F${}_{2}$) layer is destabilized 
when the current magnitude is larger than a critical value, 
\begin{equation}
  j_{\rm c}
  =
  \frac{2 \alpha eM_{2}d_{2}}{\hbar \vartheta \mathcal{P}}
  H_{\rm K},
  \label{eq:critical_current}
\end{equation}
where $\mathcal{P}$ is given by 
\begin{equation}
\begin{split}
  \mathcal{P}
  =&
  \frac{g_{{\rm r}({\rm F}_{k}/{\rm N})} g_{{\rm F}_{k}/{\rm N}}^{\prime}}{2 \tilde{g}_{k} [g_{{\rm r}({\rm F}_{k}/{\rm N})} + g_{{\rm F}_{k}/{\rm N}}^{\prime}]}
\\
  &
  \times
  \left[
    \frac{2 (1 - \lambda_{1}^{\prime} \lambda_{2}^{\prime})}{(1-\lambda_{1}^{\prime}\lambda_{2}^{\prime} m_{1z}^{2})^{2}}
    +
    \frac{\lambda_{1}^{\prime}}{1-\lambda_{1}^{\prime} \lambda_{2}^{\prime} m_{1z}^{2}}
  \right]
  m_{1y}
  m_{1z}
\end{split}
\end{equation}

% =========================================================================================== %
% =========================================================================================== %

\begin{figure}
  \centerline{\includegraphics[width=1.0\columnwidth]{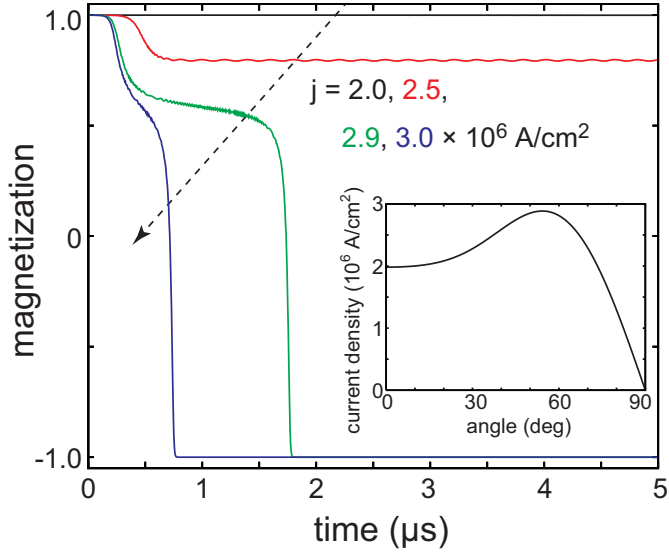}}
  \caption{
           The time evolution of $m_{2z}$ for $j=2.0$, $2.5$, $2.9$, and $3.0 \times 10^{6}$ A/cm${}^{2}$. 
           The inset shows the dependence of the balance current density $j(E)$ on 
           $\theta=\cos^{-1}m_{z}=\cos^{-1}\sqrt{-2E/(MH_{\rm K})}$. 
  \vspace{-3.5ex}}
  \label{fig:fig2}
\end{figure}

% =========================================================================================== %
% =========================================================================================== %

We confirm the deterministic switching of $\mathbf{m}_{2}$ from $\mathbf{m}_{2}=+\mathbf{e}_{z}$ to $\mathbf{m}_{2}=-\mathbf{e}_{z}$ 
by numerically solving Eq. (\ref{eq:LLG}). %the LLG equation, Eq. (\ref{eq:LLG}). 
The values of parameters are taken from typical experiments in CoFeB/TaN heterostructure [16] 
or similar metallic F/N multilayer [22-26] as 
$\lambda_{\rm F}=12$ nm, 
$\rho_{\rm F}=1600$ $\Omega$nm, 
$\beta=0.56$, 
$d_{\rm F}=1$ nm, 
$\lambda_{\rm N}=2.5$ nm,
$\rho_{\rm N}=3750$ $\Omega$nm,
$d_{\rm N}=4$ nm, 
$\vartheta=0.1$, 
$r=0.28$ k$\Omega$nm${}^{2}$,
$\gamma=0.70$, 
$g_{\rm r}/S=15$ nm${}^{-2}$, 
$\gamma_{0}=1.764 \times 10^{7}$ rad/(Oe s),
$\alpha=0.005$,
$M_{2}=1250$ emu/cm${}^{3}$,
and $H_{\rm K}=450$ Oe, 
where, we assume that the interface conductances, $g$ and $g_{\rm r}$, at the F${}_{k}$/N${}_{k}$ and F${}_{k}$/N interfaces are the same. 
We assume that the magnetization of the F${}_{1}$ layer points to the direction 
$\mathbf{m}_{1}=(\sin\theta_{\rm p}\cos\varphi_{\rm p},\sin\theta_{\rm p}\sin\varphi_{\rm p},\cos\theta_{\rm p})$ 
where $\theta_{\rm p}=30^{\circ}$ and $\varphi_{\rm p}=90^{\circ}$. 
The critical current density estimated by Eq. (\ref{eq:critical_current}) is $1.98 \times 10^{6}$ A/cm${}^{2}$. 
Figure \ref{fig:fig2} shows the time evolution of $m_{2z}$ 
obtained from Eq. (\ref{eq:LLG}) with these parameters and several current densities. 
The dynamics becomes relatively slow around $m_{2z}\sim 0.6-0.8$ where $\mathbf{m}_{2}$ becomes almost parallel to $\mathbf{m}_{1z}$, 
and therefore, the second and third torques in Eq. (\ref{eq:LLG_STT}) become small. 
As shown, the deterministic switching from the positive to negative $z$-direction is confirmed. 

% =========================================================================================== %

The switching occurs at the current density of $j_{\rm sw}\simeq 2.9 \times 10^{6}$ A/cm${}^{2}$, 
which is larger than the critical current density. 
Note that the critical current determines the instability of the initial state, 
while the instability does not guarantee the switching. 
Let us define the balance current density $j(E)$ as a current satisfying 
$\oint {\rm d}t ({\rm d}E/{\rm d}t)=0$, %=\oint {\rm d}t[-M \mathbf{H}_{2}\cdot({\rm d}\mathbf{m}_{2}/{\rm d}t)]=0$, 
where $E=-M \int {\rm d}\mathbf{m}_{2}\cdot\mathbf{H}_{2}=-MH_{\rm K}m_{2z}^{2}/2$ is the energy density, 
and the integral region is over the constant energy curve. 
The constant energy curve corresponds to 
a precession trajectory with a constant angle $\theta=\cos^{-1}m_{z}=\cos^{-1}\sqrt{-2E/(MH_{\rm K})}$. 
As discussed in Ref. [31], 
the critical and switching current densities are defined as 
$j_{\rm c}=\lim_{E \to E_{\rm min}}j(E)$ and $j_{\rm sw}={\rm max}[j(E)]$, respectively, 
which are not necessarily same. 
The analytical solution of $j(E)$ for an arbitrary $E$ is complex. 
Instead, we show the numerically calculated $j(E)$ as a function 
of $\theta$ in the inset of Fig. \ref{fig:fig2}. 
As shown, $j(E)$ has a maximum at $\theta \simeq 55^{\circ}$. 
Therefore, $j(\theta \simeq 55^{\circ})=2.9 \times 10^{6}$ A/cm${}^{2}$ is the theoretical switching current density, 
which is consistent with the numerical result in Fig. \ref{fig:fig2}. 
On the other hand, $j_{\rm c}$ corresponds to $j(\theta \to 0^{\circ})$. 
When the current density is larger than $j_{\rm c}$ and smaller than $j_{\rm sw}$, 
the self-oscillation of the magnetization can be expected, 
which is beyond the scope of this paper. 

% =========================================================================================== %

\section{Conclusion}
\label{sec:Conclusion}

In conclusion, 
we propose a model of deterministic magnetization switching 
in a perpendicular ferromagnet by spin Hall effect. 
The system consists of two ferromagnets and three nonmagnets. 
The nonmagnets generate pure spin current by the spin Hall effect, 
whose spin polarization points to the in-plane ($y$) direction. 
Passing through a ferromagnet however, the direction of the spin polarization is modified 
due to the different relaxation length scales of the longitudinal and transverse spin currents. 
Consequently, a tilt of the magnetization in one ferromagnet breaks the symmetry of the system, 
and the spin Hall torque excited in such geometry %by a pure spin current flowing through a such ferromagnet 
enables to switch a perpendicular magnetization in another ferromagnet deterministically. 
The idea was confirmed by deriving the spin Hall torque formula 
%from the diffusion equation of the spin accumulation and the Landauer formula, 
and solving the LLG equation numerically. %with the spin Hall torque numerically. 

% =========================================================================================== %

% =========================================================================================== %

\section*{Acknowledgment}

T. T. expresses gratitude to 
Sinji Yuasa, Kay Yakushiji, Koji Ando, Hitoshi Kubota, Akio Fukushima, Takayuki Nozaki, Makoto Konoto, 
Hidekazu Saito, Satoshi Iba, Aurelie Spiesser, Yoichi Shiota, Sumito Tsunegi, Ryo Hiramatsu, Takehiko Yorozu, 
Hiroki Maehara, and Ai Emura for their support and encouragement. 
%S. Yuasa, K. Yakushiji, K. Ando, H. Kubota, A. Fukushima, T. Nozaki, M. Konoto, 
%H. Saito, S. Iba, A. Spiesser, Y. Shiota, S. Tsunegi, R. Hiramatsu, T. Yorozu, 
%H. Maehara, and A. Emura for their support and encouragement. 
This work was supported by JSPS KAKENHI Grant-in-Aid for Young Scientists (B) 25790044. 

% =========================================================================================== %

\ifCLASSOPTIONcaptionsoff
  \newpage
\fi

%\bibliography{biblist}% Produces the bibliography via BibTeX.

% =========================================================================================== %

\end{document}